\documentclass[preprint, aps, prd]{revtex4-1}

\def\pd{\partial}
\def\mc{\mathcal}

\usepackage[dvips]{graphicx}
\usepackage{amssymb}
\usepackage{amssymb,amsmath}
\usepackage{graphicx}
\makeatother
\begin{document}

\title{Gravity duals of 5D N=2 SYM from F(4) gauged
supergravity}

\author{Parinya Karndumri}
\email[REVTeX Support: ]{parinya.ka@hotmail.com} \affiliation{String
Theory and Supergravity Group, Department of Physics, Faculty of
Science, Chulalongkorn University, 254 Phayathai Road, Pathumwan,
Bangkok 10330, Thailand}

\date{\today}
\begin{abstract}
We study gravity duals of the minimal $N=2$ super Yang-Mills (SYM)
gauge theories in five dimensions using the matter coupled $F(4)$
gauged supergravity in six dimensions. The $F(4)$ gauged
supergravity coupled to $n$ vector multiplets contains $4n+1$ scalar
fields, parametrized by $\mathbb{R}^+\times SO(4,n)/SO(4)\times
SO(n)$ coset manifold. Maximally supersymmetric vacua of the gauged
supergravity with $SU(2)\times G$ gauge group, with $G$ being an
$n$-dimensional subgroup of $SO(n)$, correspond to five dimensional
superconformal field theories (SCFTs) with $SU(2)_R$ R-symmetry and
$G$ global symmetry. Deformations of the UV SCFTs for $G=SU(2)$ and
$G=U(2)\sim SU(2)\times U(1)$ symmetries that lead to non-conformal
$N=2$ SYM with various unbroken global symmetries are studied
holographically.
\end{abstract}
\maketitle

\section{Introduction}
Much insight to strongly coupled gauge theories can be gained from
studying their gravity duals via the AdS/CFT correspondence
\cite{maldacena} and its generalization to non-conformal field
theories \cite{DW/QFT_townsend,correlator_DW/QFT,Skenderis_DW/QFT}.
One consequence of the AdS/CFT correspondence which has been
extensively studied is holographic RG flows. These flows describe
deformations of a UV conformal field theory (CFT) to another
conformal fixed point or to a non-conformal field theory in the IR.
On the gravity side, an RG flow in the dual field theory is
described by an asymptotically AdS solution which becomes AdS space
in certain limit corresponding to the UV CFT. The gravity solutions
interpolate between this AdS space and another AdS space in the case
of flows to some IR fixed points. For flows to non-conformal field
theories, gravity solutions in the IR will take the form of a domain
wall \cite{non_CFT_flow}. Furthermore, in flows between CFTs, bulk
scalar fields take finite constant values at both conformal fixed
points while in flows to non-conformal theories, they are usually
logarithmically divergent.
\\
\indent The above argument leads to gravity duals of various
supersymmetric gauge theories in four dimensions, and many important
characteristics of the gauge theories such as gaugino condensates
and confinements can be successfully described by gravity solutions
of five dimensional gauged supergravity, see for example
\cite{Dual_of_4DSYM1,Dual_of_4DSYM2,Dual_of_4DSYM3}. On the other
hand, holographic duals of higher dimensional gauge theories have
not much been explored in the literatures. In this paper, we will
carry out a similar study for $N=2$ supersymmetric Yang-Mills (SYM)
gauge theories in five dimensions using six-dimensional $F(4)$
gauged supergravity. This should provide the 5-dimensional analogue
of the 4-dimensional results in
\cite{Dual_of_4DSYM1,Dual_of_4DSYM2,Dual_of_4DSYM3}.
\\
\indent Five-dimensional field theories are interesting in their own
right. It has been discovered in \cite{Seiberg_5Dfield,
Seiberg_5Dfield2,Seiberg_5Dfield3} that five-dimensional gauge
theories admit non-trivial fixed points with enhanced global
symmetry. The 5D field theory describes the dynamics of the D4/D8
brane system whose near horizon limit gives rise to $AdS_6$ geometry
\cite{D4D8}. At the fixed points, the $SO(2N_f)\times U(1)$ global
symmetry of the gauge theory with $N_f<8$ flavors is enhanced to
$E_{N_f+1}$. $E_{6,7,8}$ are the usual exceptional groups and other
groups are defined by $E_1=SU(2)$, $E_2=SU(2)\times U(1)$,
$E_3=SU(3)\times SU(2)$, $E_4=SU(5)$ and $E_5=SO(10)$
\cite{Seiberg_5Dfield}. This symmetry enhancement in the case of
$SU(2)$ gauge theories has also been shown to appear in the
superconformal indices \cite{5DSymmetry_enhanced}.
\\
\indent By using AdS$_6$/CFT$_5$ correspondence, it has been
proposed in \cite{ferrara_AdS6} that five-dimensional superconformal
field theories with global symmetry $G$ should correspond to $AdS_6$
vacua of the matter coupled $F(4)$ gauged supergravity in the
six-dimensional bulk with $SU(2)_R\times G$ gauge group. The
$SU(2)_R$ R-symmetry is gauged by three of the four vector fields in
the supergravity multiplet while the $G$ part of the gauge group is
gauged by the vectors in the vector multiplets. The dual field
theory has been identified with a singleton field theory on the
boundary. A number of papers on gauge/gravity correspondence
involving 5D gauge theories and the generalization to quiver gauge
theories from the ten-dimensional point of view have appeared in
\cite{Bergman,Bergman2,Bergman3}. RG flows between 5D quiver gauge
theories with $N_f=0$ have been studied recently in \cite{5D_flow}
in the ten-dimensional context. Holographic RG flows within in the
framework of $F(4)$ gauged supergravity have also been studied in
\cite{F4_nunezAdS6} and \cite{F4_flow}. In this paper, we will give
another example of flow solutions to 5D non-conformal gauge theories
in the framework of six-dimensional gauged supergravity. As in lower
dimensions, this should be more convenient to work with than the
ten-dimensional computation and could provide a useful tool in the
holographic study of $N=2$ 5D SYM.
\\
\indent Furthermore, the study of gravity duals of 5D gauge theories
is not only important in AdS$_6$/CFT$_5$ correspondence but is also
useful in the context of AdS$_7$/CFT$_6$ correspondence
\cite{Douglas_5Dgauge_theory,5Dtheory_inAdS7_CFT6}. This originates
from the proposal that the less understood $N=(2,0)$ gauge theory in
six dimensions could be defined in term of 5D SYM. Furthermore, it
has been shown that 5D SCFT could be an IR fixed point of $N=2^*$
gauge theory in four dimensions \cite{4D_5D_flow}. Therefore, having
gravity duals of 5D SYM could be very useful in understanding the
dynamics of M5-branes and gauge theories in other dimensions as
well.
\\
\indent The paper is organized as follow. In section \ref{F4_SUGRA},
we review relevant information about matter coupled $F(4)$ gauged
supergravity in six dimensions and formulae used throughout the
paper. Holographic RG flows to non-conformal field theories from the
UV fixed point identified with the maximally supersymmetric $AdS_6$
critical points will be given in section \ref{E1} and \ref{E2}. All
of the solutions can be analytically obtained and would be more
useful than the numerical solutions given in some other cases. We
end the paper by giving some conclusions and comments in section
\ref{conclusion}.
\section{Matter coupled $F(4)$ gauged supergravity and the dual $N=2$ Super Yang-Mills}\label{F4_SUGRA}
We begin with a brief review of the matter coupled $F(4)$ gauged
supergravity in six dimensions. The theory is an extension of the
pure $F(4)$ gauged supergravity, constructed long time ago in
\cite{F4_Romans}, by coupling $n$ vector multiplets to the $N=(1,1)$
supergravity multiplet. The resulting theory is elegantly
constructed by using the superspace approach in \cite{F4SUGRA1,
F4SUGRA2, F4SUGRA3}. In the present work, we will need only
supersymmetry transformations of fermions and the bosonic Lagrangian
involving the metric and scalars. Most of the notations and
conventions are the same as those given in \cite{F4SUGRA1} and
\cite{F4SUGRA2} but with the metric signature $(-+++++)$.
\\
\indent In half-maximal $N=(1,1)$ supersymmetry, the field content
of the supergravity multiplet is given by
\begin{displaymath}
\left(e^a_\mu,\psi^A_\mu, A^\alpha_\mu, B_{\mu\nu}, \chi^A,
\sigma\right)
\end{displaymath}
where $e^a_\mu$, $\chi^A$ and $\psi^A_\mu$ denote the graviton, the
spin-$\frac{1}{2}$ field and the gravitini, respectively.
Both$\chi^A$ and $\psi^A_\mu$ are eight-component pseudo-Majorana
spinors with indices $A,B=1,2$ referring to the fundamental
representation of the $SU(2)_R\sim USp(2)_R$ R-symmetry. The
remaining fields are given by the dilaton $\sigma$, four vectors
$A^\alpha_\mu,\, \alpha=0,1,2,3$, and a two-form field $B_{\mu\nu}$.
\\
\indent A vector multiplet has component fields
\begin{displaymath}
(A_\mu,\lambda_A,\phi^\alpha).
\end{displaymath}
Each multiplet will be labeled by an index $I=1,\ldots, n$. The $4n$
scalars $\phi^{\alpha I}$ are described by a symmetric quaternionic
manifold $SO(4,n)/SO(4)\times SO(n)$. The dilaton $\sigma$ can also
be regarded as living in the coset space $\mathbb{R}^+\sim O(1,1)$.
As in \cite{F4SUGRA1}, it is convenient to decompose the $\alpha$
index into $\alpha=(0,r)$ in which $r=1,2,3$. The $SU(2)_R$
R-symmetry is identified with the diagonal subgroup of $SU(2)\times
SU(2)\sim SO(4)\subset SO(4)\times SO(n)$. A general compact gauge
group is then given by $SU(2)\times G$ with $\textrm{dim}\, G=n$.
\\
\indent The $4n$ scalars living in the $SO(4,n)/SO(4)\times SO(n)$
coset can be parametrized by the coset representative
$L^\Lambda_{\phantom{as}\Sigma}$, $\Lambda,\Sigma=0,\ldots , 3+n$.
Using the index splitting $\alpha=(0,r)$, we can split
$L^\Lambda_{\phantom{as}\Sigma}$ into
$(L^\Lambda_{\phantom{as}\alpha},L^\Lambda_{\phantom{as}I})$ and
further to $(L^\Lambda_{\phantom{as}0}, L^\Lambda_{\phantom{as}r},
L^\Lambda_{\phantom{as}I})$. The vielbein of the
$SO(4,n)/SO(4)\times SO(n)$ coset $P^I_\alpha$ can be obtained from
the left-invariant 1-form of $SO(4,n)$
\begin{equation}
\Omega^\Lambda_{\phantom{sa}\Sigma}=
(L^{-1})^\Lambda_{\phantom{sa}\Pi}\nabla L^\Pi_{\phantom{sa}\Sigma},
\qquad \nabla
L^\Lambda_{\phantom{sa}\Sigma}=dL^\Lambda_{\phantom{sa}\Sigma}
-f^{\phantom{s}\Lambda}_{\Gamma\phantom{sa}\Pi}A^\Gamma
L^\Pi_{\phantom{sa}\Sigma},
\end{equation}
via
\begin{equation}
P^I_\alpha=(P^I_{\phantom{sa}0},P^I_{\phantom{sa}r})=(\Omega^I_{\phantom{sa}0},\Omega^I_{\phantom{sa}r}).
\end{equation}
The structure constants of the full gauge group $SU(2)_R\times G$
are denoted by $f^\Lambda_{\phantom{as}\Pi\Sigma}$ which can be
split into $\epsilon_{rst}$ and $C_{IJK}$ for $SU(2)_R$ and $G$,
respectively. The direct product structure of the gauge group
$SU(2)_R\times G$ leads to two coupling constants $g_1$ and $g_2$
which, in the above equation, are encoded in
$f^\Lambda_{\phantom{as}\Pi\Sigma}$.
\\
\indent In this paper, we are interested in $n=3,4$ cases with gauge
groups $SU(2)_R\times SU(2)$ and $SU(2)_R\times SU(2)\times U(1)$.
To describe $SO(4,n)/SO(4)\times SO(n)$, we introduce basis elements
of $(4+n)\times (4+n)$ matrices by
\begin{equation}
(e^{xy})_{zw}=\delta_{xz}\delta_{yw}, \qquad w,x,y,z=1,\ldots, n+4\,
.
\end{equation}
The $SO(4)$, $SU(2)_R$ and non-compact generators of $SO(4,n)$ are
accordingly given by
\begin{eqnarray}
SO(4)&:&\qquad
J^{\alpha\beta}=e^{\beta+1,\alpha+1}-e^{\alpha+1,\beta+1},\qquad \alpha,\beta=0,1,2,3,\nonumber \\
SU(2)_{R}&:&\qquad J^{rs}=e^{s+1,r+1}-e^{r+1,s+1},\qquad r,s=1,2,3,\nonumber \\
& &Y^{\alpha I}=e^{\alpha+1,I+4}+e^{I+4,\alpha+1},\qquad I=1,\ldots,
n\, .
\end{eqnarray}
\indent Gaugings lead to fermionic mass-like terms and the scalar
potential in the Lagrangian as well as some modifications to the
supersymmetry transformations at first order in the coupling
constants. We will give only information relevant to the study of
supersymmetric RG flows and refer the reader to \cite{F4SUGRA1} and
\cite{F4SUGRA2} for more details and complete formulae. The bosonic
Lagrangian for the metric and scalar fields is given by
\cite{F4SUGRA2}
\begin{equation}
\mathcal{L}=\frac{1}{4}eR-e\pd_\mu \sigma\pd^\mu \sigma
-\frac{1}{4}eP_{I\alpha\mu}P^{I\alpha\mu}-eV
\end{equation}
where $e=\sqrt{-g}$. The scalar kinetic term is written in term of
$P^{I\alpha}_\mu=P^{I\alpha}_i\pd_\mu\phi^i$, $i=1,\ldots, 4n$. For
completeness, we also give the explicit form of the scalar potential
\begin{eqnarray}
V&=&-e^{2\sigma}\left[\frac{1}{36}A^2+\frac{1}{4}B^iB_i-\frac{1}{4}\left(C^I_{\phantom{s}t}C_{It}+4D^I_{\phantom{s}t}D_{It}\right)\right]
-m^2e^{-6\sigma}\mc{N}_{00}\nonumber \\
& &+me^{-2\sigma}\left[\frac{2}{3}AL_{00}-2B^iL_{0i}\right]
\end{eqnarray}
where $\mc{N}_{00}$ is the $00$ component of the scalar matrix
defined by
\begin{equation}
\mc{N}_{\Lambda\Sigma}=L^{\phantom{as}0}_\Lambda
(L^{-1})_{0\Sigma}+L^{\phantom{as}i}_\Lambda
(L^{-1})_{i\Sigma}-L^{\phantom{as}I}_\Lambda (L^{-1})_{I\Sigma}\, .
\end{equation}
Various quantities appearing in the scalar potential and in the
supersymmetry transformations given below are defined as follow
\begin{eqnarray}
A&=&\epsilon^{rst}K_{rst},\qquad B^i=\epsilon^{ijk}K_{jk0},\\
C^{\phantom{ts}t}_I&=&\epsilon^{trs}K_{rIs},\qquad D_{It}=K_{0It}
\end{eqnarray}
where
\begin{eqnarray}
K_{rst}&=&g_1\epsilon_{lmn}L^l_{\phantom{r}r}(L^{-1})_s^{\phantom{s}m}L_{\phantom{s}t}^n+
g_2C_{IJK}L^I_{\phantom{r}r}(L^{-1})_s^{\phantom{s}J}L_{\phantom{s}t}^K,\nonumber
\\
K_{rs0}&=&g_1\epsilon_{lmn}L^l_{\phantom{r}r}(L^{-1})_s^{\phantom{s}m}L_{\phantom{s}0}^n+
g_2C_{IJK}L^I_{\phantom{r}r}(L^{-1})_s^{\phantom{s}J}L_{\phantom{s}0}^K,\nonumber
\\
K_{rIt}&=&g_1\epsilon_{lmn}L^l_{\phantom{r}r}(L^{-1})_I^{\phantom{s}m}L_{\phantom{s}t}^n+
g_2C_{IJK}L^I_{\phantom{r}r}(L^{-1})_I^{\phantom{s}J}L_{\phantom{s}t}^K,\nonumber
\\
K_{0It}&=&g_1\epsilon_{lmn}L^l_{\phantom{r}0}(L^{-1})_I^{\phantom{s}m}L_{\phantom{s}t}^n+
g_2C_{IJK}L^I_{\phantom{r}0}(L^{-1})_I^{\phantom{s}J}L_{\phantom{s}t}^K\,
.
\end{eqnarray}
\indent Finally, the supersymmetry transformations of $\chi^A$,
$\lambda^I_A$ and $\psi^A_\mu$ involving only scalars and the metric
are given by
\begin{eqnarray}
\delta\psi_{\mu
A}&=&D_\mu\epsilon_A-\frac{1}{24}\left(Ae^\sigma+6me^{-3\sigma}(L^{-1})_{00}\right)\epsilon_{AB}\gamma_\mu\epsilon^B\nonumber
\\
& &-\frac{1}{8}
\left(B_te^\sigma-2me^{-3\sigma}(L^{-1})_{t0}\right)\gamma^7\sigma^t_{AB}\gamma_\mu\epsilon^B,\label{delta_psi}\\
\delta\chi_A&=&\frac{1}{2}\gamma^\mu\pd_\mu\sigma\epsilon_{AB}\epsilon^B+\frac{1}{24}
\left[Ae^\sigma-18me^{-3\sigma}(L^{-1})_{00}\right]\epsilon_{AB}\epsilon^B\nonumber
\\
& &-\frac{1}{8}
\left[B_te^\sigma+6me^{-3\sigma}(L^{-1})_{t0}\right]\gamma^7\sigma^t_{AB}\epsilon^B\label{delta_chi}\\
\delta
\lambda^{I}_A&=&P^I_{ri}\gamma^\mu\pd_\mu\phi^i\sigma^{r}_{\phantom{s}AB}\epsilon^B+P^I_{0i}
\gamma^7\gamma^\mu\pd_\mu\phi^i\epsilon_{AB}\epsilon^B-\left(2i\gamma^7D^I_{\phantom{s}t}+C^I_{\phantom{s}t}\right)
e^\sigma\sigma^t_{AB}\epsilon^B \nonumber
\\
& &-2me^{-3\sigma}(L^{-1})^I_{\phantom{ss}0}
\gamma^7\epsilon_{AB}\epsilon^B\label{delta_lambda}
\end{eqnarray}
where $\sigma^{tC}_{\phantom{sd}B}$ are Pauli matrices, and
$\epsilon_{AB}=-\epsilon_{BA}$. The space-time gamma matrices
$\gamma^a$, with $a$ being tangent space indices, satisfy
\begin{equation}
\{\gamma^a,\gamma^b\}=2\eta^{ab},\qquad
\eta^{ab}=\textrm{diag}(-1,1,1,1,1,1),
\end{equation}
and $\gamma^7=\gamma^0\gamma^1\gamma^2\gamma^3\gamma^4\gamma^5$.
\\
\indent We now give a short description of the UV SCFT which is
identified with the $AdS_6$ vacuum preserving 16 supercharges. At
this vacuum, all scalars vanish, and the full gauge group
$SU(2)_R\times G$ is preserved. The bulk fields in the supergravity
multiplet are dual to the operators in the the energy-momentum
tensor supermultiplet in the five-dimensional field theory while the
bulk vector multiplets correspond to the global current
supermultiplets. The full spectrum of all supergravity fields can be
found in \cite{F4SUGRA1} and \cite{F4SUGRA2}. $SU(2)_R$ singlet
scalars in the adjoint representation of $G$ are dual to operators
of dimension 4 corresponding to the highest components of the global
current supermultiplets. These scalars give supersymmetry preserving
deformations as discussed in \cite{ferrara_AdS6}. On the other hand,
the dilaton and $SU(2)_R$ triplet scalars are dual to operators of
dimension 3 and correspond to supersymmetry breaking deformations.

\section{RG flows from $SU(2)_R\times SU(2)$ SCFT}\label{E1}
We begin with the simplest possibility with $n=3$ and $SU(2)_R\times
SU(2)$ gauge group. The gravity theory consists of 13 scalars
parametrized by $O(1,1)\times SO(4,3)/SO(4)\times SO(3)$ coset
space. We are interested in $SU(2)_R$ singlet scalars which are
given by $\sigma$ and additional 3 scalars from $SO(4,3)/SO(4)\times
SO(3)$. The latter correspond to the non-compact generators
$Y_{11}$, $Y_{12}$ and $Y_{13}$. The coset representative is
accordingly written as
\begin{equation}
L=e^{a_1Y_{11}}e^{a_2Y_{12}}e^{a_3Y_{13}}\, .
\end{equation}
The space-time metric is the standard domain wall ansatz
\begin{equation}
ds^2=e^{2A(r)}dx^2_{1,4}+dr^2
\end{equation}
in which five-dimensional Poincare symmetry is manifest. From now
on, the six-dimensional space-time indices will be split as
$(\mu,r)$ with $\mu=0,\ldots, 4$.
\\
\indent Using \eqref{delta_psi}, \eqref{delta_chi} and
\eqref{delta_lambda}, we find the following BPS equations
\begin{eqnarray}
a_1'&=&-2e^{-3\sigma}m \frac{\sinh{a_1}}{\cosh{a_2}\cosh{a_3}},\label{E1_eq1}\\
a_2'&=&-2e^{-3\sigma}m\frac{\cosh{a_1}\sinh{a_2}}{\cosh{b_3}},\label{E1_eq2}\\
a_3'&=&-2e^{-3\sigma}m\cosh{a_1}\cosh{a_2}\sinh{a_3},\label{E1_eq3}\\
\sigma'&=&-\frac{1}{2}\left[e^\sigma g_1-3e^{-3\sigma}m\cosh{a_1}\cosh{a_2}\cosh{a_3}\right],\label{E1_eq4}\\
A'&=&\frac{1}{2}\left[e^{\sigma}g_1+e^{-3\sigma}m\cosh{a_1}\cosh{a_2}\cosh{a_3}\right]\label{E1_eq5}
\end{eqnarray}
where $'$ denotes $\frac{d}{dr}$ and we have used the projection
$\gamma^r\epsilon^A=\epsilon^A$. The presence of $\gamma^7$ in
$\delta\lambda^I_A$ does not impose any condition on $\epsilon^A$
since it appears as an overall factor in all of the BPS equations
obtained from $\delta\lambda^I_A=0$. That the bulk gravity solution
preserves eight supercharges is to be expected because the minimal
SYM in five dimensions has 8 supercharges. The equation for the warp
factor $A(r)$ is obtained from $\delta \psi^A_\mu$, $\mu
=0,1,2,3,4$. The $\delta\psi_r^A=0$ equation would give the
dependence of the Killing spinors on the $r$-coordinate as in other
cases. We now look at solutions of interest.
\subsection{Flow to $SU(2)_R\times U(1)$ SYM}
We first study the solution that breaks the $SU(2)$ global symmetry
to $U(1)$. This corresponds to turning on only $a_3$ and $\sigma$.
The latter is of course a singlet of the full gauge group
$SU(2)_R\times SU(2)$. With $a_1=a_2=0$, equations \eqref{E1_eq1}
and \eqref{E1_eq2} are trivially satisfied, and equations
\eqref{E1_eq3}, \eqref{E1_eq4} and \eqref{E1_eq5} become
\begin{eqnarray}
 a_3'&=&-2e^{-3\sigma}m\sinh{a_3},\label{E1_U1_eq1}\\
 \sigma'&=&\frac{1}{2}\left(-g_1e^\sigma+3e^{-3\sigma}m\cosh{a_3}\right),\label{E1_U1_eq2}\\
 A'&=&\frac{1}{2}\left(g_1e^\sigma+e^{-3\sigma}m\cosh{a_3}\right).\label{E1_U1_eq3}
\end{eqnarray}
We can solve equation \eqref{E1_U1_eq1} by introducing a new radial
coordinate $\tilde{r}$ such that
$\frac{d\tilde{r}}{dr}=e^{-3\sigma}$. We then find the solution for
$a_3$
\begin{equation}
a_3=\pm\ln
\left[\frac{1+e^{-2m\tilde{r}+C_1}}{1-e^{-2m\tilde{r}+C_1}}\right]\,
.
\end{equation}
This form is very similar to the solution studied in
\cite{Dual_of_4DSYM1} for the 4D SYM. $C_1$ is an integration
constant. There are two possibilities for the two signs. Combining
equations \eqref{E1_U1_eq1} and \eqref{E1_U1_eq2} gives an equation
for $\frac{d\sigma}{da_3}$
\begin{equation}
\frac{d\sigma}{da_3}=\frac{1}{4m}\left(e^{4\sigma}g_1\textrm{csch}a_3-3m\coth
a_3\right)
\end{equation}
whose solution is given by
\begin{equation}
\sigma=-\frac{1}{4}\ln\left[ \frac{g_1\left(3\cosh a_3-\cosh
(3a_3)+18C_2\sinh^3a_3\right)} {6m}\right]
\end{equation}
with $C_2$ being another integration constant.
\\
\indent After changing to $\tilde{r}$ coordinate and using $a_3$
solution, we find that the combination
\eqref{E1_U1_eq3}+\eqref{E1_U1_eq2} becomes, with now $'$ being
$\frac{d}{d\tilde{r}}$,
\begin{equation}
A'+\sigma'=\frac{2m\left(e^{4m\tilde{r}}+e^{2C_1}\right)}{e^{2C_1}-e^{4m\tilde{r}}}\,
.
\end{equation}
The solution to this equation can be readily found to be
\begin{equation}
A=2m\tilde{r}+\ln\left(1-e^{C_1-2m\tilde{r}}\right)+\ln\left(1+e^{C_1-2m\tilde{r}}\right)-\sigma
\end{equation}
where we have neglected the additive integration constant to $A$ by
absorbing it into the rescaling of the $x^\mu$ coordinates. To
identify the maximally supersymmetric vacuum at $\sigma=a_3=0$ with
the $N=2$ SCFT, we have to set $g_1=3m$. In the above solutions, we
have not done this in order to keep the solutions in a generic form.
Note also that if we try to truncate $\sigma$ out by setting
$\sigma=0$, equation \eqref{E1_U1_eq2} will imply $a_3=0$.
Therefore, to obtain a non-trivial solution, we must keep $\sigma$
non-vanishing.
\\
\indent An RG flow to a non-conformal field theory with only the
dilaton $\sigma$ in pure $F(4)$ gauged supergravity has been studied
in \cite{F4_nunezAdS6}. The resulting solution is interpreted as the
analogue of the Coulomb branch flow. We now have a more general flow
solution in the case of matter coupled $F(4)$ gauged supergravity.
As $r\rightarrow \infty$, $\sigma,a_3\rightarrow 0$, we see that
$\tilde{r}\sim r\rightarrow \infty$. In this limit, we obtain the
maximally supersymmetric $AdS_6$ background with $A\sim 2mr
=\frac{r}{L}$ where the $AdS_6$ radius in the UV is given by
$L=\frac{1}{2m}$. According to the AdS/CFT correspondence, this is
identified with the UV SCFT with $SU(2)_R\times SU(2)$ SCFT in five
dimensions. From the above solutions, the behavior of $\sigma$ and
$a_3$ near the UV point with $g_1=3m$ is readily seen to be
\begin{equation}
a_3\sim e^{-2mr}=e^{-\frac{r}{L}},\qquad\sigma\sim a_3^3\sim
e^{-6mr}=e^{-\frac{3r}{L}} \, .
\end{equation}
We see that $a_3$ corresponds to a deformation by a relevant
operator of dimension $\Delta=4$ while $\sigma$ describes a
deformation by a vacuum expectation value of operator of dimension
$\Delta=3$.
\\
\indent There is an issue of singularities in the IR which are
typical in flows to non-conformal field theories. Physical and
unphysical singularities can be classified by using the criterion
given in \cite{Gubser_singularity}. From the solution, we see that
$a_3$ is singular when $\tilde{r}\rightarrow \frac{C_1}{2m}$. We now
consider the case with $a_3>0$ and $a_3<0$, separately. For $a_3>0$,
we find $a_3= -\ln \left(2m\tilde{r}-C_1\right)+\ln 2$ as
$2m\tilde{r}\sim C_1$ and
\begin{eqnarray}
\sigma &=&\frac{3}{4}\ln
\left(2m\tilde{r}-C_1\right)-\frac{1}{4}\ln\left[9C_2-2+3(2m\tilde{r}-C_1)^2(9C_2-2)\right.\nonumber
\\
&
&\left.+3(9C_2+2)(2m\tilde{r}-C_1)^4+\ldots\right]\label{sigma_behave1}
\end{eqnarray}
The warp factor $A$ near $\tilde{r}\rightarrow \frac{C_1}{2m}$ is
given by
\begin{equation}
A=\ln \left(2m\tilde{r}-C_1\right)-\sigma\, .
\end{equation}
For $C_2= \frac{2}{9}$, we find that
\begin{equation}
\sigma\sim -\frac{1}{4}\ln (2m\tilde{r}-C_1),\qquad A\sim
\frac{5}{4}\ln (2m\tilde{r}-C_1).
\end{equation}
We can find the relation between $\tilde{r}$ and $r$ in this limit
by using $\frac{d\tilde{r}}{dr}=e^{-3\sigma}$. The relation is given
by
\begin{equation}
2mr-C=4(2m\tilde{r}-C_1)^4
\end{equation}
where $C$ is a new integration constant. The metric becomes
\begin{equation}
ds^2=\left(2mr-C\right)^{10}dx^2_{1,4}+dr^2
\end{equation}
where we have absorbed the multiplicative constant to the scaling of
$x^\mu$ coordinates. According to the
Domain-Wall/Quantum-Field-Theory (DW/QFT) correspondence, this
background is dual to a non-conformal SYM theory in five dimensions.
\\
\indent To determine whether the singularity in the solution is
acceptable or not, we check the scalar potential on the solution.
With $a_1=a_2=0$ and $g_1=3m$, the potential is given by
\begin{equation}
V=e^{-6\sigma}m^2\left[\cosh(2a_3)-12e^{4\sigma}\cosh
a_3-9e^{8\sigma}\right].
\end{equation}
It can be verified that $V\rightarrow -\infty$ as
$a_3,\sigma\rightarrow \infty$. The singularity is then physical
according to the criterion of \cite{Gubser_singularity}. For
$a_3<0$, it can be easily checked that the singularity is acceptable
for the choice $C_2=-\frac{2}{9}$ which leads to
\begin{eqnarray}
a_3&\sim & \ln(2m\tilde{r}-C_1),\qquad \sigma
-\frac{1}{4}\ln(2m\tilde{r}-C_1),\nonumber \\
ds^2&=&\left(2mr-C\right)^{10}dx^2_{1,4}+dr^2\, .
\end{eqnarray}
\\
\indent On the other hand, if $C_2\neq \pm\frac{2}{9}$ for
$a_3\sim\pm \ln (2m\tilde{r}-C_1)$, respectively, the solution is
asymptotic to
\begin{eqnarray}
a_3 &\sim &\pm \ln (2m\tilde{r}-C_1),\qquad \sigma \sim
\frac{3}{4}\ln
(2m\tilde{r}-C_1),\nonumber \\
ds^2&=&\left(2mr-C\right)^{\frac{2}{13}}dx^2_{1,4}+dr^2\, .
\end{eqnarray}
where we have used the relation
$(2m\tilde{r}-C_1)^{\frac{13}{4}}=\frac{13}{4}\left(2mr-C\right)$,
near $\tilde{r}\sim \frac{C_1}{2m}$, with a constant $C$. The
singularity in this case is however not acceptable since
$V\rightarrow \infty$.
\\
\indent It is useful to comment on the IR singularities. Following
the discussion in \cite{non_CFT_flow}, the criterion of
\cite{Gubser_singularity} is related to the fact that the divergence
in a vacuum expectation value of an operator $O$ dual to a canonical
scalar $\phi$ is excluded. In the IR, the scalar bulk action is
given by $S\sim \int e^{5A}(\pd \phi)^2$ since the potential is
irrelevant due to the divergence of the scalar. The expectation
value of $O$ is then given by
\begin{equation}
\langle O\rangle\sim \frac{\delta S}{\delta \phi}\sim e^{5A}\pd_r
\phi\sim (r-r_0)^{5\kappa -1}
\end{equation}
where we have used the asymptotic behavior $\phi\sim \phi_0\ln
(r-r_0)$ and $A\sim \kappa \ln (r-r_0)$. The singularity occurs at
$r=r_0$. We see that $\langle O\rangle $ diverges when
$\kappa<\frac{1}{5}$. In the present case, the physical flow has
$\kappa =5$ while the unphysical one has $\kappa=\frac{1}{13}$. This
is consistent with the finiteness of the expectation value of the
dual operator.

\subsection{Flow to $SU(2)_R$ SYM}\label{E1_to_E0}
If the other scalars $a_1$ and $a_2$ are non-vanishing, the solution
will break the $SU(2)$ global symmetry, completely. It is now more
difficult to solve all five BPS equations, but it turns out that
these equations admit analytic solutions.
\\
\indent To obtain the solution, we consider $A$, $\sigma$, $a_1$ and
$a_2$ as functions of $a_3$. Combining equations \eqref{E1_eq2} and
\eqref{E1_eq3}, we find
\begin{equation}
\frac{da_2}{da_3}=\frac{\tanh a_2}{\sinh a_3\cosh a_3}\, .
\end{equation}
This is easily solved by
\begin{equation}
a_2=\ln
\left[\frac{e^{2a_3+C_1}-e^{C_1}+\sqrt{(1+e^{2a_3})^2+e^{2C_1}(e^{2a_3}-1)}}{1+e^{2a_3}}\right]
=\sinh^{-1}\left(e^{C_1}\tanh a_3\right).
\end{equation}
Similarly, by solving equations \eqref{E1_eq1} and \eqref{E1_eq3},
we obtain
\begin{equation}
a_1=\sinh^{-1}\frac{e^{C_2}\sinh
a_3}{\sqrt{1-e^{2C_1}+(1+e^{2C_1})\cosh(2a_3)}}\, .
\end{equation}
Using the $a_1$ and $a_2$ solutions and the new radial coordinate
$\tilde{r}$, we find the solution for $a_3$
\begin{equation}
a_3=\pm\frac{1}{2}\cosh^{-1}\left[\frac{e^{2C_2}+2e^{2C_1}-2+4\tanh^2(2m\tilde{r}-C_3)}{2+2e^{2C_1}+e^{2C_2}}\right].\label{E1_noSym_a3Sol}
\end{equation}
We can similarly solve for $\sigma$ as a function of $a_3$. The
solution is given by
\begin{eqnarray}
\sigma &=&\frac{1}{4}\ln\left[3 m
\left(\tilde{A}^2+\tilde{B}^2\right)^2 \text{csch}^6a_3  \left(36
\tilde{A}^2 C_4 \left(\tilde{A}^2+\tilde{B}^2\right)^2 \sinh ^3 a_3
\left(\tilde{A}^2 \cosh (2
a_3)+\tilde{B}^2\right)\right.\right.\nonumber \\
& &\left.\left.\qquad -2 \left(3 \tilde{A}^2+\tilde{B}^2-2
\tilde{A}^2 \cosh (2 a_3)\right) \left(\tilde{A}^2 \cosh (2
a_3)+\tilde{B}^2\right)^{3/2}\right)
\right] \nonumber \\
& &-\frac{1}{4}\ln \left[1296 \tilde{A}^4 C_4^2 g_1
\left(\tilde{A}^2+\tilde{B}^2\right)^4 \left(\tilde{A}^2 \cosh (2
a_3)+\tilde{B}^2\right)\right.\nonumber \\
& &\left.-4 g_1 \text{csch}^6 a_3 \left(\tilde{A}^4 \cosh (4
a_3)+\tilde{A}^4+\tilde{A}^2 \left(\tilde{B}^2-3 \tilde{A}^2\right)
\cosh (2 a_3)-3 \tilde{A}^2
\tilde{B}^2-\tilde{B}^4\right)^2\right].\nonumber \\
\end{eqnarray}
We have defined two new constants
$\tilde{A}=\sqrt{2+2e^{2C_1}+e^{2C_2}}$ and
$\tilde{B}=\sqrt{2-2e^{2C_1}-e^{2C_2}}$ for conveniences.
\\
\indent Finally, adding \eqref{E1_eq4} to \eqref{E1_eq5} and
changing the variable from $r$ to $a_3$, we find a simple equation
for $A$
\begin{equation}
\frac{dA}{da_3}+\frac{d\sigma}{da_3}=-\coth a_3
\end{equation}
whose solution is
\begin{equation}
A=-\sigma-\ln (\sinh a_3).
\end{equation}
\indent Near the UV point, we find $r\sim \tilde{r}\rightarrow
\infty$, $a_1\sim a_2\sim a_3\sim e^{-\frac{r}{L}}$ and $\sigma\sim
e^{-\frac{3r}{L}}$. The solution for $A$ then gives $A\sim
2mr=\frac{r}{L}$. The flow is again driven by turning on operators
of dimension four corresponding to $a_{1,2,3}$ and a vev of a
dimension three operator dual to $\sigma$.
\\
\indent It can be checked by expanding \eqref{E1_noSym_a3Sol} that
$a_3\rightarrow \pm\infty$ as $2m\tilde{r}\rightarrow \tilde{C}$
where we have collectively denoted all constant terms from the
expansion by $\tilde{C}$. The behavior of $a_3$ near this point is
$a_3\sim \pm\ln(2m\tilde{r}-\tilde{C}) $. Although $a_3$ blows up
when $2m\tilde{r}\sim \tilde{C}$, $a_1$ and $a_2$ remain finite with
$a_2\sim \sinh^{-1}e^{C_1}$ and
$a_1\sim\sinh^{-1}\frac{e^{C_2}}{\sqrt{2+2e^{2C_1}}}$. Similar to
the previous case, the criterion of \cite{Gubser_singularity}
requires
$C_4=\pm\frac{2\sqrt{2}\tilde{A}}{9(\tilde{A}^2+\tilde{B}^2)^2}$ for
the singularity to be physical. This is true for both $a_3<0$ and
$a_3>0$. We find that
\begin{eqnarray}
a_3 &\sim &\pm\ln(2m\tilde{r}-\tilde{C}),\qquad \sigma \sim
-\frac{1}{4}\ln(2m\tilde{r}-\tilde{C}),\nonumber \\
ds^2 &=&(2mr-C)^{10}dx^2_{4,1}+dr^2\, .
\end{eqnarray}
It can be readily verified that there always exist the values of
$C_1$ and $C_2$ at which this behavior gives $V\rightarrow -\infty$.
\\
\indent For $C_4\neq
\pm\frac{2\sqrt{2}\tilde{A}}{9(\tilde{A}^2+\tilde{B}^2)^2}$, the
solution near $2m\tilde{r}\sim \tilde{C}$ becomes
\begin{eqnarray}
a_3 &\sim &\pm\ln(2m\tilde{r}-\tilde{C}),\qquad  \sigma\sim
-\frac{3}{4}a_3=\frac{3}{4}\ln (2m\tilde{r}-\tilde{C}),\nonumber \\
 ds^2&=&(2mr-C)^{\frac{2}{13}}dx^2_{1,4}+dr^2\, .
\end{eqnarray}
This solution is not physical as can be checked that $V\rightarrow
\infty$ for all values of $C_1$ and $C_2$.

\subsection{Flow to $SU(2)_\textrm{diag}$ SYM}
In this subsection, we will look at an RG flow with
$SU(2)_{\textrm{diag}}\sim (SU(2)_R\times SU(2))_{\textrm{diag}}$
singlet scalars. Some non-supersymmetric $AdS_6$ vacua and
holographic RG flows interpolating between these critical points and
the maximally supersymmetric $AdS_6$ have been studied in
\cite{F4_flow}. In this work, we will give a supersymmetric flow to
a non-conformal field theory.
\\
\indent There is only one singlet scalar under
$SU(2)_{\textrm{diag}}$ from $\frac{SO(4,3)}{SO(4)\times SO(3)}$,
see the detail in \cite{F4_flow}. The coset representative can be
written as
\begin{equation}
L=e^{a(Y_{21}+Y_{32}+Y_{43})}\, .
\end{equation}
The supersymmetry transformations of $\psi^A_\mu$, $\chi^A$ and
$\lambda^I_A$ give the following BPS equations
\begin{eqnarray}
a'&=&-e^{\sigma}\sinh(2a)\left(g_1\cosh a-g_2 \sinh a\right),\label{E1_SU2d_eq1}\\
\sigma'&=&\frac{1}{2}e^{-3\sigma}\left[3m+e^{4\sigma}\left(g_2\sinh^3 a-g_1\cosh^3a\right)\right],\label{E1_SU2d_eq2}\\
A'&=&\frac{1}{2}e^{-3\sigma}\left[m+e^{4\sigma}\left(g_1\cosh^3
a-g_2\sinh^3a\right)\right].\label{E1_SU2d_eq3}
\end{eqnarray}
 Note that for non-singlet scalars of $SU(2)_R$, the $SU(2)$ coupling $g_2$ appears.
 \\
 \indent In order to solve the above equations, we will treat $\sigma$ and $A$ as functions of $a$
\begin{equation}
\frac{d\sigma}{da}=\frac{3me^{-4\sigma}-g_1\cosh^3a+g_2\sinh^3a}{2\sinh(2a)\left(g_1\cosh
a-g_2\sinh a\right)}
\end{equation}
which can be solved by
\begin{equation}
\sigma =\frac{1}{4}\ln\left[ \frac{6m\cosh (2a)+C_1\sinh
(2a)}{2g_1\cosh a-2g_2\sinh a}\right].\label{sigma_sol3}
\end{equation}
We can check that as $a\rightarrow 0$ and $g_1=3m$,
$\sigma\rightarrow 0$ as expected for the UV point. This is the case
for any value of $C_1$. To solve for $a$ from equation
\eqref{E1_SU2d_eq1}, it is convenient to define a new coordinate
$\tilde{r}$ via $e^\sigma=\frac{d\tilde{r}}{dr}$. Only in this case,
$\tilde{r}$ is defined by $e^\sigma=\frac{d\tilde{r}}{dr}$. In all
other cases, we always have $e^{-3\sigma}=\frac{d\tilde{r}}{dr}$.
\\
\indent With this new variable, we can solve for $\tilde{r}$ as a
function of $a$. The resulting solution is given by
\begin{eqnarray}
2g_1g_2\tilde{r}&=&g_2\ln
\coth\frac{a}{2}-2g_1\tan^{-1}\left[\tanh\frac{a}{2}\right]\nonumber
\\
& &+2\sqrt{g_1^2-g_2^2}\tan^{-1}\left[\frac{g_1\tanh\frac{a}{2}-g_2}
{\sqrt{g_1^2-g_2^2}}\right]\label{SU2D_a1_sol}
\end{eqnarray}
where we have neglected the additive integration constant.
\\
\indent Taking the combination \eqref{E1_SU2d_eq2}-$3\times$
\eqref{E1_SU2d_eq3} with \eqref{E1_SU2d_eq1}, we can rewrite
equation for $A$ as
\begin{equation}
\frac{d\sigma}{da}-3\frac{dA}{da}=\frac{g_1\sinh a+g_2(1-\cosh
a)}{g_1\cosh a-g_2\sinh a}\, .
\end{equation}
The solution is readily obtained to be
\begin{equation}
A=\frac{1}{3}\left[\sigma+\ln \sinh (2a)+\ln (g_1\cosh a-g_2\sinh
a)\right].
\end{equation}
From the above solutions, we can find the behavior of $a$, $\sigma$
and $A$ near the UV point, $a=\sigma=0$. In this limit,
$\tilde{r}\sim r\rightarrow \infty$, we find $a\sim \sigma \sim
e^{-6mr}=e^{-\frac{3r}{L}}$ and $A\sim 2mr=\frac{r}{L}$. This
indicates that the flow is driven by vacuum expectation values of
operators of dimension three. This is to be expected since it has
been pointed out in \cite{F4_flow} that the flow driven by turning
on the operators dual to $\sigma$ and $a$ corresponds to a
non-supersymmetric flow to a non-supersymmetric IR fixed point. In
the IR, there are a number of possibilities depending on the values
of $g_2$ and the integration constant $C_1$ since these lead to
different IR behaviors of $a$ and $\sigma$.
\\
\indent We begin with the $g_2=g_1$ case and consider the solution
for large $|a|$. For $a<0$, we find by expanding the solution in
\eqref{SU2D_a1_sol} that $a$ diverges as $a\sim \frac{1}{3}\ln
(g_1\tilde{r}-\tilde{C})$. As in the previous case, we have
collectively denoted all of the constants by $\tilde{C}$. When
$C_1=6m$, the solutions for $\sigma$ and $A$ become
\begin{eqnarray}
\sigma &\sim &
\frac{1}{4}\ln(g_1\tilde{r}-\tilde{C}),\qquad A\sim \frac{7}{36}\ln (g_1\tilde{r}-\tilde{C}),\nonumber \\
ds^2 &=&(3mr-C)^{\frac{14}{27}}dx^2_{1,4}+dr^2\, .
\end{eqnarray}
This leads to $V\rightarrow -\infty$ which is acceptable.
\\
\indent For $C_1\neq 6m$, we find different behavior
\begin{eqnarray}
\sigma &\sim &
-\frac{1}{12}\ln(g_1\tilde{r}-\tilde{C}),\qquad A\sim \frac{1}{12}\ln (g_1\tilde{r}-\tilde{C}),\nonumber \\
ds^2 &=&(g_1r-C)^{\frac{2}{13}}dx^2_{1,4}+dr^2
\end{eqnarray}
which gives $V\rightarrow \infty$ as expected since in this case
$\kappa <\frac{2}{5}$.
\\
\indent For $a>0$, we find that $a\sim -\ln
(g_1\tilde{r}-\tilde{C})$. There are two possibilities for $C_1=-6m$
and $C_1\neq -6m$ which give respectively
\begin{eqnarray}
\sigma &\sim &
\frac{1}{4}\ln(g_1\tilde{r}-\tilde{C}),\qquad A\sim \frac{13}{12}\ln (g_1\tilde{r}-\tilde{C}),\nonumber \\
ds^2 &=&(g_1r-C)^{\frac{26}{9}}dx^2_{1,4}+dr^2
\end{eqnarray}
and
\begin{eqnarray}
\sigma &\sim &
-\frac{3}{4}\ln(g_1\tilde{r}-\tilde{C}),\qquad A\sim \frac{3}{4}\ln (g_1\tilde{r}-\tilde{C}),\nonumber \\
ds^2 &=&(g_1r-C)^{\frac{6}{7}}dx^2_{1,4}+dr^2\, .
\end{eqnarray}
Both of them give $V\rightarrow -\infty$. We then conclude that for
$g_2=g_1$, all flows with $a>0$ are physical, but flows with $a<0$
are physical only for $C_1=6m$.
\\
\indent We now move to the $g_1\neq g_2$ case and quickly look at
$a>0$ and $a<0$ flows, separately. With $a>0$, the solution becomes
\begin{eqnarray}
a&\sim & -\frac{1}{3}\ln \left[(g_1-g_2)\tilde{r}-\tilde{C}\right],\qquad \sigma \sim -\frac{1}{12}\ln \left[(g_1-g_2)\tilde{r}-\tilde{C}\right],\nonumber \\
ds^2 &=&
\left[(g_1-g_2)\tilde{r}-\tilde{C}\right]^{\frac{2}{13}}dx^2_{1,4}+dr^2,
\end{eqnarray}
for $C_1\neq -6m$, and
\begin{eqnarray}
a&\sim & -\frac{1}{3}\ln \left[(g_1-g_2)\tilde{r}-\tilde{C}\right],\qquad \sigma \sim \frac{1}{4}\ln \left[(g_1-g_2)\tilde{r}-\tilde{C}\right],\nonumber \\
ds^2 &=&
\left[(g_1-g_2)\tilde{r}-\tilde{C}\right]^{\frac{14}{27}}dx^2_{1,4}+dr^2,
\end{eqnarray}
for $C_1=-6m$. The former is unphysical, but the latter is physical
provided that $-(5+4\sqrt{2})m<g_2< (4\sqrt{2}-5)m$.
\\
\indent Finally, for $a<0$, we find the IR behavior
\begin{eqnarray}
a&\sim & \frac{1}{3}\ln \left[(g_1+g_2)\tilde{r}-\tilde{C}\right],\qquad \sigma \sim -\frac{1}{12}\ln \left[(g_1+g_2)\tilde{r}-\tilde{C}\right],\nonumber \\
ds^2 &=&
\left[(g_1+g_2)\tilde{r}-\tilde{C}\right]^{\frac{2}{13}}dx^2_{1,4}+dr^2,
\end{eqnarray}
for $C_1\neq 6m$, and
\begin{eqnarray}
a&\sim & \frac{1}{3}\ln \left[(g_1+g_2)\tilde{r}-\tilde{C}\right],\qquad \sigma \sim \frac{1}{4}\ln \left[(g_1+g_2)\tilde{r}-\tilde{C}\right],\nonumber \\
ds^2 &=&
\left[(g_1+g_2)\tilde{r}-\tilde{C}\right]^{\frac{14}{27}}dx^2_{1,4}+dr^2,
\end{eqnarray}
for $C_1=6m$. Similar to the previous case, only the second
possibility is physical provided that
$(5-4\sqrt{2})m<g_2<(5+4\sqrt{2})m$. In summary, for $g_2\neq g_1$,
flows with $a>0$ and $a<0$ are physical for $C_1=-6m$ and $C_1=6m$,
respectively for some appropriate values of $g_2$.

\section{RG flows from $SU(2)_R\times U(2)$ SCFT}\label{E2}
To give more examples, we consider $F(4)$ gauged supergravity
coupled to four vector multiplets with $SU(2)_R\times SU(2)\times
U(1)$ gauge group. There are 16 scalars parametrized by
$SO(4,4)/SO(4)\times SO(4)$ coset. We will focus on $SU(2)_R$
singlet scalars which are highest components of the global symmetry
multiplet and correspond to supersymmetry preserving deformations.
Together with the dilaton $\sigma$, there are five $SU(2)_R$ singlet
scalars. The coset representative can be written as
\begin{equation}
L=e^{a_1Y_{11}}e^{a_2Y_{12}}e^{a_3Y_{13}}e^{a_4Y_{14}}\, .
\end{equation}
Using the projector $\gamma^r\epsilon^A=\epsilon^A$, we can derive
the following BPS equations
\begin{eqnarray}
a_1'&=&-\frac{2me^{-3\sigma}\sinh a_1}{\cosh a_2\cosh a_3\cosh
a_4},\label{E2_eq1}\\
a_2'&=&-\frac{2me^{-3\sigma}\sinh a_2\cosh a_1}{\cosh a_3\cosh
a_4},\label{E2_eq2}\\
a_3'&=&-\frac{2me^{-3\sigma}\cosh a_1\cosh a_2\sinh a_3}{\cosh
a_4},\label{E2_eq3}\\
a_4'&=&-2me^{-3\sigma}\cosh a_1\cosh a_2\cosh a_3\sinh
a_4,\label{E2_eq4}\\
\sigma'&=&\frac{1}{2}\left[3me^{-3\sigma}\cosh a_1\cosh a_2\cosh
a_3\cosh a_4-g_1e^{\sigma}\right],\label{E2_eq5}\\
A'&=&\frac{1}{2}\left[me^{-3\sigma}\cosh a_1\cosh a_2\cosh a_3\cosh
a_4+g_1e^{\sigma}\right].\label{E2_eq6}
\end{eqnarray}
\indent We are interested in the RG flows with the symmetry breaking
patterns $U(2)\rightarrow SU(2)$, $U(2)\rightarrow U(1)\times U(1)$,
$U(2)\rightarrow U(1)$ and the completely broken $U(2)$. The
procedure is essentially the same as in the previous section, so we
will neglect some details and simply give the solutions.

\subsection{Flow to $SU(2)_R\times SU(2)$ SYM}
In order to preserve $SU(2)\subset SU(2)\times U(1)$, only $a_4$ is
allowed to be non-vanishing. The above equations reduce to three
simple equations
\begin{eqnarray}
a_4'&=&-2me^{-3\sigma}\sinh a_4,\label{E2_SU2_eq1}\\
\sigma'&=&\frac{1}{2}\left(3me^{-3\sigma}\cosh
a_4-g_1e^{\sigma}\right),\label{E2_SU2_eq2}\\
A'&=&\frac{1}{2}\left(me^{-3\sigma}\cosh
a_4+g_1e^{\sigma}\right).\label{E2_SU2_eq3}
\end{eqnarray}
By introducing a new radial coordinate $\tilde{r}$ via
$\frac{d\tilde{r}}{dr}=e^{-3\sigma}$ as in the previous section, we
find the solutions
\begin{eqnarray}
a_4&=&\pm\ln
\left[\frac{1+e^{-2m\tilde{r}+C_1}}{1-e^{-2m\tilde{r}+C_1}}\right],\nonumber
\\
\sigma &=&-\frac{1}{4}\ln\left[ \frac{g_1\left(3\cosh a_4-\cosh
(3a_4)+18C_2\sinh^3a_4\right)} {6m}\right] ,\nonumber
\\
A&=&2m\tilde{r}+\ln\left(1-e^{C_1-2m\tilde{r}}\right)+\ln\left(1+e^{C_1-2m\tilde{r}}\right)-\sigma.
\end{eqnarray}
\indent Near the UV point, $a_4$, $\sigma$ and $A$ behave as
\begin{equation}
a_4\sim e^{-2mr}, \qquad \sigma\sim e^{-6mr},\qquad A\sim 2mr\, .
\end{equation}
\indent Similar to the previous solutions, we find that the IR
singularity at $\tilde{r}\sim \frac{C_1}{2m}$ is physical for
$a_4\sim \pm \ln (2m\tilde{r}-C_1)$ if we choose
$C_2=\pm\frac{2}{9}$. In both cases, the IR metric is given by
\begin{equation}
ds^2=\left(2mr-C\right)^{10}dx^2_{1,4}+dr^2\, .
\end{equation}
Other choices of $C_2$ lead to unacceptable singularities.

\subsection{Flow to $SU(2)_R\times U(1)\times U(1)$ SYM}
In this subsection, we will give the solution for the flow to SYM
with $SU(2)_R\times U(1)^2$ symmetry. To find this solution, we set
$a_1=a_2=a_4=0$. The BPS equations, which are similar to the
previous subsection, give the following solutions, in term of
$\tilde{r}$ coordinate,
\begin{eqnarray}
a_3&=&\pm\ln
\left[\frac{1+e^{-2m\tilde{r}+C_1}}{1-e^{-2m\tilde{r}+C_1}}\right],\nonumber
\\
\sigma &=&-\frac{1}{4}\ln\left[ \frac{g_1\left(3\cosh a_3-\cosh
(3a_3)+18C_2\sinh^3a_3\right)} {6m}\right] ,\nonumber
\\
A&=&2m\tilde{r}+\ln\left(1-e^{C_1-2m\tilde{r}}\right)+\ln\left(1+e^{C_1-2m\tilde{r}}\right)-\sigma.
\end{eqnarray}
Near the UV point, we find $a_3\sim e^{-2mr}$, $\sigma\sim e^{-6mr}$
and $A\sim 2mr$. In the IR, $\tilde{r}\rightarrow \frac{C_1}{m}$,
the physical solution with $C_2=\pm\frac{2}{9}$ is given by
\begin{eqnarray}
a_4 &\sim &\pm \ln (2m\tilde{r}-C_1),\qquad \sigma \sim
-\frac{1}{4}\ln (2m\tilde{r}-C_1),\nonumber \\
ds^2&=&\left(2mr-C\right)^{10}dx^2_{1,4}+dr^2\, .
\end{eqnarray}

\subsection{Flow to $SU(2)_R\times U(1)$ SYM}
We then consider the flow that breaks $SU(2)\times U(1)$ global
symmetry to $U(1)$. In this case, we turn on both $a_3$ and $a_4$.
This leads to more complicated equations due to the coupling between
$a_4$ and $a_3$. We will regard $a_4$ as a new variable and find
that the solutions for $a_3$, $\sigma$ and $A$ are given by
\begin{eqnarray}
a_3&=&\sinh^{-1}\left[e^{C_1}\tanh a_4\right],\nonumber \\
\sigma &=&-\frac{1}{4}\ln
\left[\frac{g_1}{6\sqrt{2}m}\left[72C_2\sinh^3 a_4(1+e^{2C_1})\right.\right.\nonumber \\
& &\left.\left.\phantom{\frac{1}{2}}-2\cosh a_4
\left[(1+e^{2C_1})\cosh(2a_4)-e^{2C_1}-2\right]\sqrt{2+2e^{2C_1}\tanh^2a_4}\right]\right]
,\nonumber \\
A&=&-\sigma-\ln\sinh a_4\, .
\end{eqnarray}
The solution of $a_4$ in term of $\tilde{r}$ is given by
\begin{equation}
\tilde{r}=\frac{1}{2m}\tanh^{-1}\sqrt{\frac{1+\cosh(2a_4)+2e^{2C_1}\sinh^2a_4}{2}}\,
.
\end{equation}
\indent At the UV point, we find the expected behavior $a_{3,4}\sim
e^{-2mr}$, $\sigma \sim e^{-6mr}$ and $A\sim 2mr$. In the IR, we
consider the behavior of the solutions as $a_4\rightarrow \infty$.
In this limit, the $a_4$ solution becomes $a_4\sim -\ln
(2m\tilde{r}-\tilde{C})$ for some constant $\tilde{C}$. We find that
the requirement for the IR singularity to be acceptable is given by
$C_2=\frac{1}{9}\sqrt{\frac{1+e^{2C_1}}{2}}$. The behavior of $a_3$,
$\sigma$ and $A$ is given by
\begin{equation}
a_3\sim \sinh^{-1}e^{C_1},\qquad \sigma\sim -\frac{1}{4}\ln
(2m\tilde{r}-\tilde{C}), \qquad A\sim\frac{5}{4}\ln
(2m\tilde{r}-\tilde{C}).
\end{equation}
With the relation $2mr-C=4(2m\tilde{r}-\tilde{C})^{\frac{1}{4}}$,
the metric in the IR then takes the form of a domain wall
\begin{equation}
ds^2=(2mr-C)^{10}dx^2_{1,4}+dr^2\, .
\end{equation}

\subsection{Flow to $SU(2)_R$ SYM}\label{E2_to_E0}
We now quickly look at the flow breaking the $U(2)$
symmetry,completely. Finding the solution in this case amounts to
solving all of the six BPS equations. This however turns out to be
not difficult. The resulting solutions for $a_i$, $\sigma$ and $A$
are given by
\begin{eqnarray}
a_3 &=&\sinh^{-1}\left(e^{C_1}\tanh a_4\right),\nonumber \\
a_2&=&\sinh^{-1}\frac{e^{C_2}\sinh
a_4}{\sqrt{1-e^{2C_1}+(1+e^{2C_1})\cosh(2a_4)}},\nonumber \\
a_1&=&\sinh^{-1}\frac{e^{C_3}\sinh
a_4}{\sqrt{2-2e^{2C_1}-e^{2C_2}+(2+2e^{2C_1}+e^{2C_2})\cosh (2a_4)}},\nonumber \\
\sigma&=&\frac{1}{4}\ln
\left[96\sqrt{2}m\sqrt{4+\alpha^2-\alpha^2\textrm{sech}^2a_4}\right]\nonumber
\\
& &-\frac{1}{4}\ln\left[g_1 \left(2304 \left(\alpha^2+4\right) C_4
\sinh ^3 a_4 \sqrt{\alpha^2+4-\alpha^2
\text{sech}^2a_4}\right.\right.\nonumber \\
& & -\sqrt{2} \text{sech}a_4 \left(3
\alpha^4+\left(\alpha^2+4\right)^2 \cosh (4 a_4)+16 \alpha^2 \right.
\nonumber \\
& & \left.\left.\left.-4 \left(\alpha^4+6 \alpha^2+8\right) \cosh (2
a_4)-48\right)\right)
\right],\nonumber \\
A&=&-\sigma-\ln \sinh a_4,\nonumber \\
a_4&=&\frac{1}{2}\cosh^{-1}\left[\frac{8\tanh^2(2m\tilde{r}-C_5)+\alpha^2-4}{\alpha^2+4}\right]
\end{eqnarray}
where $\alpha=\sqrt{4e^{2C_1}+2e^{2C_2}+e^{2C_3}}$. At the UV fixed
point, the solutions become
\begin{equation}
a_{1,2,3,4}\sim e^{-2mr},\qquad \sigma\sim e^{-6mr},\qquad A\sim
2mr\, .
\end{equation}
\indent In the IR, we have to set
$C_4=\frac{1}{144}\sqrt{\frac{4+\alpha^2}{2}}$ in order to obtain a
physical solution. The solution is then given by
\begin{eqnarray}
& &a_4\sim -\ln (2m\tilde{r}-\tilde{C}),\qquad a_3\sim \sinh^{-1}e^{C_1},\nonumber \\
& &a_2\sim \sinh^{-1}\frac{e^{C_2}}{\sqrt{2+2e^{2C_1}}},\qquad
a_1\sim \sinh^{-1}\frac{e^{C_3}}{\sqrt{4+4e^{2C_1}+2e^{2C_2}}}
,\nonumber \\
& &\sigma \sim -\frac{1}{4}\ln (2m\tilde{r}-\tilde{C}),\qquad A\sim
\frac{5}{4}\ln (2m\tilde{r}-\tilde{C}),\nonumber \\
& &ds^2=(2mr-C)^{10}dx^2_{1,4}+dr^2\, .
\end{eqnarray}
\indent All of the flows given above are driven by turning on
operators of dimension 4 and a vev of a dimension 3 operator.

\section{Conclusions}\label{conclusion}
We have studied various holographic RG flows from matter coupled
$F(4)$ gauged supergravity. These flows describe deformations of the
UV $N=2$ SCFTs with $SU(2)$ and $SU(2)\times U(1)$ global symmetries
in five dimensions to non-conformal $N=2$ SYM theories in the IR. We
have explored various symmetry breaking patterns and interpreted the
solutions as RG flows driven by turning on operators of dimension 4
in a vacuum with non-zero vev of a dimension 3 operator dual to the
six-dimensional dilaton except for the flow to
$SU(2)_{\textrm{diag}}$ SYM which is driven by vacuum expectation
values of dimension three operators. We have also identified
physical flows which have acceptable IR singularities from the
resulting solutions. Therefore, these solutions might be useful in
the study of strongly coupled $N=2$ SYM in five dimensions. However,
the identification of the dual five-dimensional SYM corresponding to
these solutions in the IR is not clear. Accordingly, the precise
physical interpretation of these solutions needs to be clarified.
\\
\indent It is interesting to holographically compute various
characteristics of the 5D gauge theories such as the Wilson loops as
done in \cite{5D_Wilson_loop}. It could be useful to do this
computation with the six-dimensional solutions given here similar to
the four dimensional gauge theories studied in
\cite{Dual_of_4DSYM1,Dual_of_4DSYM2}. The solutions found in this
paper would hopefully be useful in this aspect and other holographic
calculations. It will very interesting (if possible) to find a
gravity solution describing the enhancement of the global symmetry
$SO(2N_f)\times U(1)$ to the $E_{N_f+1}$ fixed point in five
dimensions. In this aspect, the six-dimensional framework considered
here may not be able to accommodate this solution since the symmetry
enhancement is not seen at the classical supergravity level as
remarked in \cite{Bergman3}.
\\
\indent  It is presently not known how to embed the six-dimensional
$F(4)$ gauged supergravity coupled to $n$ vector multiplets to 10 or
11 dimensions although the pure $F(4)$ gauged supergravity and the
theory coupled to $20$ vector multiplets are known to originate from
massive type IIA compactification on warped $S^4$ and $K3$,
respectively \cite{Massive_IIA_onS4,Massive_IIA_onK3}. The embedding
of $F(4)$ gauged supergravity in type IIB theory via the non-abelian
T-duality has been proposed recently in \cite{Eoin_IIB_F4}. This
might also provide another mean to embed the six-dimensional gauged
supergravity in higher dimensions. It would be interesting to find
such an embedding which in turn can be used to uplift the solutions
found here and in \cite{F4_flow} to ten dimensions. This could
provide some insight to the dynamics of D4/D8-brane system. We hope
to come back to these issues in future works.


\begin{acknowledgments}
This work is partially supported by Chulalongkorn University through
Ratchadapisek Sompote Endowment Fund under grant GDNS57-003-23-002
and The Thailand Research Fund (TRF) under grant TRG5680010.
\end{acknowledgments}



\begin{thebibliography}{99}
\bibitem{maldacena} J. M. Maldacena, ``The large $N$ limit of
superconformal field theories and supergravity'', Adv. Theor. Math.
Phys. \textbf{2} (1998) 231-252, arXiv: hep-th/9711200.
\bibitem{DW/QFT_townsend} H.J. Boonstra, K. Skenderis and P.K.
Townsend, ``The domain-wall/QFT correspondence'', JHEP 01 (1999)
\textbf{003}, arXiv: hep-th/9807137.
\bibitem{correlator_DW/QFT} T. Gherghetta and Y. Oz, ``Supergravity, Non-Conformal Field Theories and
Brane-Worlds'', Phys. Rev. \textbf{D65} (2002) 046001, arXiv:
hep-th/0106255.
\bibitem{Skenderis_DW/QFT} Ingmar Kanitscheider, Kostas Skenderis and Marika
Taylor, ``Precision holography for non-conformal branes'', JHEP 09
(2008) \textbf{094}, arXiv: 0807.3324.
\bibitem{non_CFT_flow} M. Petrini and A. Zaffaroni, ``The holographic RG flow to conformal and non-conformal theory'', arXiv: hep-th/0002172.
\bibitem{Dual_of_4DSYM1} L. Girardello, M. Petrini, M. Porrati and A. Zaffaroni, ``The supergravity dual of $N=1$ Super Yang-Mills theory'',
Nucl. Phys. \textbf{B569} (2000) 451-469, arXiv: hep-th/9909047.
\bibitem{Dual_of_4DSYM2} L. Girardello, M. Petrini, M. Porrati and A. Zaffaroni, ``Confinement and condensates without fine tuning in supergravity duals of
gauge theories'', JHEP 05 (1999) \textbf{026}, arXiv:
hep-th/9903026.
\bibitem{Dual_of_4DSYM3} A. Khavaev and N. P. Warner, ``A class of $N=1$ supersymmetric RG flows from five-dimensional $N=8$ supergravity'',
Phys. Lett. \textbf{B495} (2000) 215-222, arXiv: hep-th/0009159.
\bibitem{Seiberg_5Dfield} N. Seiberg, ``Five dimensional SUSY field theories, non-trivial fixed points and string
dynamics'', Phys. Lett. \textbf{B388} (1996) 753-760, arXiv:
hep-th/9608111.
\bibitem{Seiberg_5Dfield2} K. Intriligator, D. R. Morrison and N.
Seiberg, ``Five dimensional supersymmetric gauge theories adn
degenerations of Calabi-Yau spaces'', Nucl. Phys. \textbf{B497}
(1997) 56, arXiv: hep-th/9702198.
\bibitem{Seiberg_5Dfield3} D. R. Morrison and N. Seiberg, ``Extremal transitions and five-dimensional supersymmetric field
theories'', Nucl. Phys. \textbf{B483} (1997) 229, arXiv:
hep-th/9609070.
\bibitem{D4D8} A. Brandhuber and Y. Oz, ``The D4-D8 brane system and five dimensional fixed
points'', Phys. Lett. \textbf{B460} (1999) 307-312, arXiv:
hep-th/9905148.
\bibitem{5DSymmetry_enhanced} D. Bashkirov, ``A comment on the enhancement of global symmetries in superconformal $SU(2)$ gauge theories
in 5D'', arXiv: 1211.4886.
\bibitem{ferrara_AdS6} S. Ferrara, A. Kehagias, H. Partouche, A.
Zaffaroni, ``AdS$_6$ interpretation of 5d superconformal field
theories'', Phys. Lett. \textbf{B431} (1998) 57-62, arXiv:
hep-th/9804006.
\bibitem{Bergman} O. Bergman and D. Rodriguez-Gomez, ``5d quivers and their AdS(6) duals'' JHEP 07 (2012) \textbf{171}, arXiv: 1206.3503.
\bibitem{Bergman2} O. Bergman and D. Rodriguez-Gomez, ``Probing the Higgs branch of 5D fixed point theories with dual
giant gravitons in AdS(6)'', JHEP 12 (2012) \textbf{047}, arXiv:
1210.0589.
\bibitem{Bergman3} O. Bergman, D. Rodriguez-Gomez and G. Zafrir, ``5D superconformal indices at large $N$ and holography'',
JHEP 08 (2013) \textbf{081}, arXiv: 1305.6870.
\bibitem{F4_nunezAdS6} U. Gursoy, C. Nunez and M. Schvellinger, ``RG flows from Spin(7), CY 4-fold
and HK manifolds to AdS, Penrose limits and pp waves'', JHEP 06
(2002) \textbf{015}, arXiv: hep-th/0203124.
\bibitem{F4_flow} P. Karndumri, ``Holographic RG flows in six dimensional F(4) gauged
supergravity'', JHEP 01 (2013) \textbf{134}, arXiv: 1210.8064.
\bibitem{5D_flow} A. Pini and D. Rodriguez-Gomez, ``Gauge/Gravity duality and RG flows in 5D gauge theories'',
arXiv: 1402.6155.
\bibitem{Douglas_5Dgauge_theory} M. R. Douglas, ``On $D=5$ super Yang-Mills theory and $(2,0)$ theory'', JHEP
02 (2011) \textbf{011}, arXiv: 1012.2880.
\bibitem{5Dtheory_inAdS7_CFT6} J. A. Minahan, A. Nedelin and M. Zabzine, ``5D super Yang-Mills theory and the correspondence to
AdS$_7$/CFT$_6$'', J. Phys. A: Math. theor. \textbf{46} (2013)
355401.
\bibitem{4D_5D_flow} C. Hoyos-Badajoz, ``Higher dimensional conformal field theories in the Coulonb branch'', Phys. Lett.
\textbf{B696} (2011) 145-150.
\bibitem{F4_Romans} L. J. Romans, ``The $F(4)$ gauged supergravity in
six-dimensions'', Nucl. Phys \textbf{B269} (1986) 691.
\bibitem{F4SUGRA1} R. D' Auria, S. Ferrara and S. Vaula, ``Matter coupled $F(4)$ supergravity and the AdS$_6$/CFT$_5$
correspondence'', JHEP 10 (2000) \textbf{013}, arXiv:
hep-th/0006107.
\bibitem{F4SUGRA2} L. Andrianopoli, R. D' Auria and S. Vaula, ``Matter coupled $F(4)$ gauged supergravity
Lagrangian'', JHEP 05 (2001) \textbf{065}, arXiv: hep-th/0104155.
\bibitem{F4SUGRA3} R. D' Auria, S. Ferrara and S. Vaula, ``AdS$_6$/CFT$_5$ correspondence for $F(4)$ gauged supergravity'',
Fortsch. Phys. \textbf{49} (2001) 459-468, arXiv: hep-th/0101066.
\bibitem{Gubser_singularity} S. S. Gubser, ``Curvature singularities: the good, the bad and the naked'', Adv. Theor.
Math. Phys. \textbf{4} (2000) 679-745.
\bibitem{5D_Wilson_loop} B. Assel, J. Estes and M. Yamazaki ``Wilson loops in 5D $N=1$ SCFTs and AdS/CFT'',
Ann. Henri Poincare \textbf{15} (2014) 589.
\bibitem{Massive_IIA_onS4} M. Cvetic, H. Lu and C. N. Pope, ``Gauged six-dimensional supergravity from
massive type IIA'', Phys. Rev. Lett. \textbf{83} (1999) 5226-5229,
arXiv: hep-th/9906221.
\bibitem{Massive_IIA_onK3} M. Haack, J. Louis and H. Singh, ``Massive type IIA theory on $K3$'',
JHEP 04 (2001) \textbf{040}, arXiv: hep-th/0102110.
\bibitem{Eoin_IIB_F4} J. Jeong, O. Kelekci and E. O Colgain, ``An alternative IIB embedding of $F(4)$ gauged
supergravity'', JHEP 05 (2013) \textbf{079}, arXiv: 1302.2105.
\end{thebibliography}
\end{document}